\documentclass[12pt]{iopart}
\usepackage{graphicx}
\begin{document}

\title {Comparison of shell model results for even-even Se isotopes }
\author{P. C. Srivastava}
\address{ Department of Physics, Indian Institute of Technology,
  Roorkee - 247667, India}
\ead{pcsrifph@iitr.ernet.in}
\author{M. J. Ermamatov}
\address{Instituto de Ciencias Nucleares, Universidad
  Nacional Aut\'onoma de M\'exico, 04510 M\'exico, D.F., Mexico}
\address{Institute of Nuclear Physics, Ulughbek, Tashkent 100214, Uzbekistan}
\begin{abstract}

Comprehensive set of shell model calculations for $^{78-84}$Se isotopes have been performed with recently
derived interactions, namely JUN45 and jj44b for 
${f_{5/2}pg_{9/2}}$ space. To study the importance of the proton excitations 
across $Z=28$ shell in this region mentioned by Cheal {\it et al.} [Phys. Rev. Lett. {\bf104}, 252502 (2010)],
calculation for ${fpg_{9/2}}$ 
valence space using an ${fpg}$ effective interaction 
 with $^{48}$Ca as core and
imposing a truncation has also been performed. Comparison of the calculations with experimental data show that the predicted results of jj44b interaction are in good agreement with experimental data.

\end{abstract} 

\submitto{\PS}
\maketitle


\section{Introduction}

The nuclei near $Z = 28$ region are subject of intensive research of  both experimental
 and theoretical investigations~\cite{Padalia,Heyde13,Otsuka13}. In particular, the region of neutron-rich nuclei
around $N = 40$ to $N = 50$ shows an evolution of shell structure. The nucleon-nucleon interaction, the spin-orbit, tensor part, and three-body effect play an important
 roles in the shell evolutions. Due to tensor interactions, the nuclear mean
field undergoes variations with neutron excess. This leads to monopole migration,
which is observed in both side of 
line. As we approach towards the neutron
drip line, the neutron density becomes very diffused, which can also leads to shell
quenching. Further motivations for nuclear structure studies are for understanding
the other open questions like, how does the filling of neutron orbitals beyond
$N = 28$ around $^{48}$Ca influence the shell structure, what is the role of tensor part
of proton-neutron interaction, how does the spin-orbit splitting between $f_{5/2}$ and
$f_{7/2}$ evolve when approaching $N = 50$, and what is the specific role of particle-hole
excitations through $Z = 28$ and $N = 40$ shell gaps for the onset of deformation
below and above $^{78}$Ni.

\begin{figure}
\begin{center}
\includegraphics[width=12.5cm]{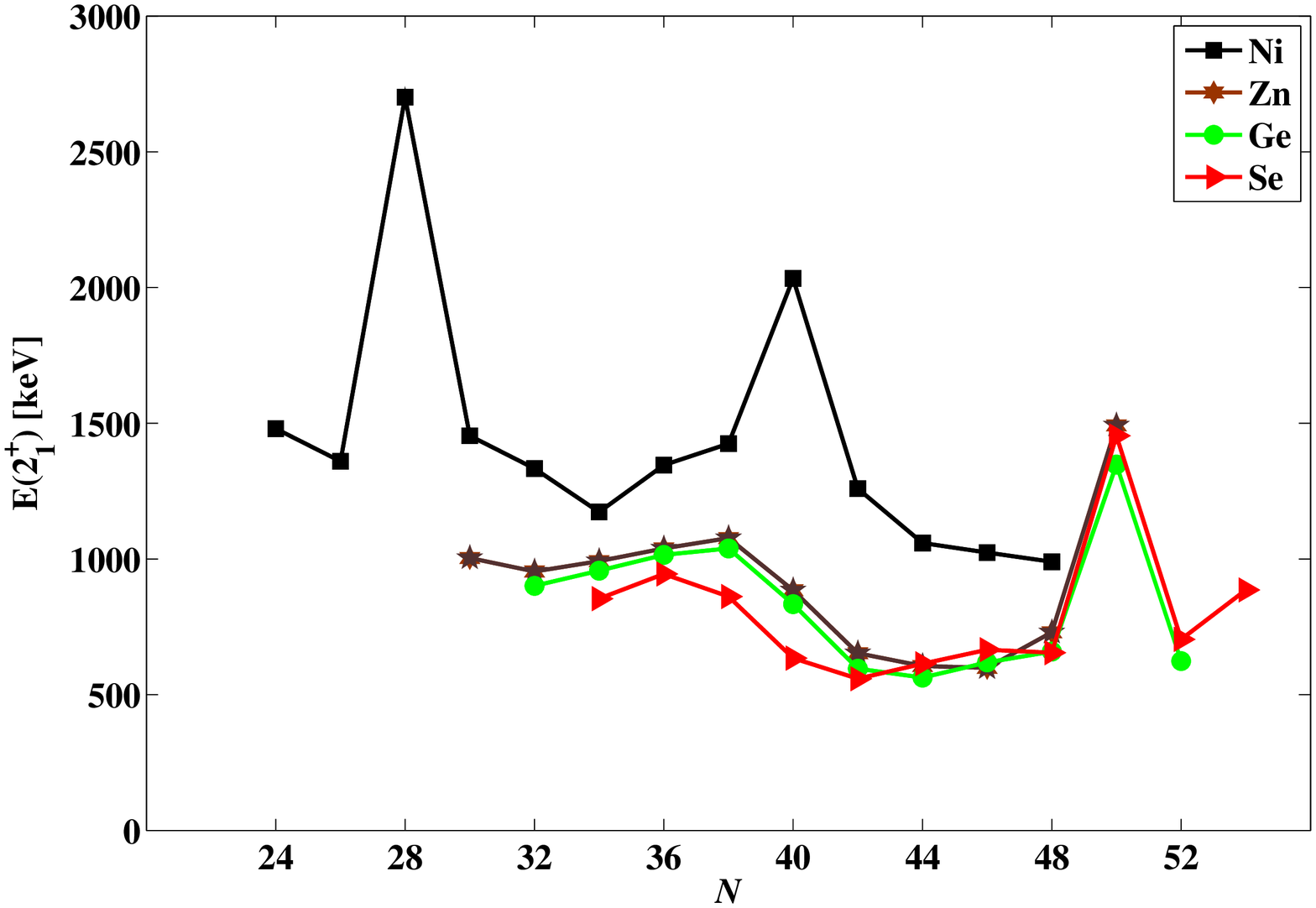}
\caption{Systematic of the experimentally observed $E(2_1^+)$ for $Z=28$ to $Z=34$ near the $N~=~28,~40$ and $~50$ shell closure.}
\label{fig1}
\end{center}
\end{figure}

To study the proton excitations across $Z=28$ gap many experimental investigations 
have been performed for Cu~\cite{flangan09}, Ga~\cite{cheal10} and As~\cite{astier11} isotopes. 
The inversion of $\pi f_{5/2}$ and $\pi p_{3/2}$ orbitals for Cu isotopes have 
been established while measuring magnetic moments for the ground-state.  
In case of $^{71-81}$Ga isotopes experimental measurement of spin and moments reveal that there is
structure change between $N=40$ and $N=50$. The $\pi f_{5/2}$ orbital dominant in the g.s. of $^{79}$Ga 
and for $^{81}$Ga, 5/2$^-$ level become ground state. The evolution of structure can be seen
from emptying of $\pi p_{3/2}$ orbital to $\pi f_{5/2}$ is started as we move from $^{71}$Ga to $^{79}$Ga.
Finally I$^\pi$=5/2$^-$ become ground state for  $^{81}$Ga. In case of $^{79}$Ga, the 
$f_{5/2}pg_{9/2}$ space fail to explain correctly the magnetic and electric quadrupole moment. 
Thus, it is crucial to include $\pi f_{7/2}$ orbital in the model space to study the effect of proton
 excitations across the $Z=28$ effect. In case of As ($Z=33$), recent experimental investigation by Astier et al
~\cite{astier11} suggests that $f_{5/2}pg_{9/2}$  space is not enough to explain quadrupole excitation built on the 
5/2$_1^-$ and 9/2$_1^+$ state of $^{81}$As.

In this work, shell model calculation for  $^{78-83}$Se isotopes in two different model spaces
$f_{5/2}pg_{9/2}$ and $fpg_{9/2}$ have been performed. The aim of this study to study the importance of proton
excitations across $Z=28$ gap. One of us recently reported importance of $\pi f_{7/2}$ orbital for predicting moments of Ga isotopes \cite{pcs_ga}. 
 In Fig.~\ref{fig1}, the systematics of $2^+$ for Ni to Se  are shown. This figure shows the persistence of $N=50$ shell closure
while $N=40$ disappears as we move from Ni to Se. 

In section~\ref{details} the details about shell model calculations is described and then spectroscopic
results for even--even $^{78,80,82,84}$Se are presented in section~\ref{E_BE}. 
In section 4,  transition probability, quadrupole moments and occupation numbers are presented.
Finally section~\ref{conc} gives the concluding remarks. 

\section{Details of model spaces and interactions}
\label{details}

The present shell model calculations have been carried out in the $f_{5/2} \,p \,g_{9/2}$ and $f \,p \,g_{9/2}$ spaces.
In the $f_{5/2} \,p \,g_{9/2}$ valence space the calculations have been performed with the interactions JUN45
\cite{Honma09} and jj44b \cite{brown}. 
  The JUN45 interaction is based on Bonn-C potential, the  single-particle energies and
 two-body matrix elements was modified empirically so as to fit 400 experimental data out of 69 nuclei with A = 63$\sim$69.  In the fitting of JUN45 interaction the experimental data are taken
around $N=50$. Thus with this interaction the shell-model results with $N \sim 50$ chain show 
reasonable agreement with experimental data.  
The jj44b interaction was obtained from a fit to about 600 binding energies and excitation energies with 30 linear combinations of the good $J-T$ two-body matrix elements. For jj44b the energy data for the fit taken from nuclei with $Z=28-30$ and $N=48-50$.  The rms deviation between experiment and theory for the the energies was 250 keV. 
The single-particle energies for the 1$p_{3/2}$,
0$f_{5/2}$, 1$p_{1/2}$ and 0$g_{9/2}$ single-particle
orbits employed
in conjunction with the JUN45 interaction are -9.8280, -8.7087, -7.8388, and -6.2617 MeV
respectively. In the case of the jj44b interaction they are -9.6566, -9.2859, -8.2695, and -5.8944
MeV, respectively. The core is $^{56}$Ni, i.e. $N = Z = 28$, and the calculations are performed in
this valence space without truncation.

In the $f \,p \,g_{9/2}$ valence space, we use a $^{48}$Ca
core, i.e. only the protons are active in the
$f_{7/2}$ orbital, and the interaction $fpg$ reported by Sorlin et al \cite{fpg}.
 The $fpg$ interaction was built using $fp$  two-body matrix
elements (TBME) from \cite{32} and $rg$ TBME 
($p_{3/2}$, $f_{5/2}$, $p_{1/2}$, and $g_{9/2}$ orbits) from \cite{33}.
For the common active orbitals in these subspaces,
matrix elements were taken from \cite{33}. As the latter
interaction ($rg$) was defined for a $^{56}$Ni core, a scaling
factor of A$^{-1/3}$ was applied to take into account the
change of radius between the $^{40}$Ca and $^{56}$Ni cores.
The remaining $f_{7/2}g_{9/2}$ TBME are taken from \cite{34}.

 The single-particle energies
are 0.0, 2.0, 4.0, 6.5 and 9.0 MeV for the 0$f_{7/2}$, 1$p_{3/2}$, 1$p_{1/2}$, 0$f_{5/2}$, and 0$g_{9/2}$ orbits, respectively.
Since the dimensionality of this valence space is prohibitively large, we have introduced a
truncation by allowing $t_\pi$ particle-hole excitations from the $\pi f_{7/2}$ orbital to the upper $fp$ orbitals
($p_{3/2}$, $f_{5/2}$, $p_{1/2}$) for protons and $t_\nu$ particle-hole excitations from the upper $fp$ orbitals to the $\nu g_{9/2}$
orbital for neutrons. The maximum allowed value for $t_\pi$ and
$t_\nu$ is four.  The maximal dimension 165000813 ( 
$\sim10^{8}$ ) is reached for positive parity in the case of $^{78}$Se when using
$f_{5/2}pg_{9/2}$ space with $^{56}$Ni core since neutron number is
furthest from the closed shell for this nucleus among the Se isotopes
considered in this work. In case of $^{78}$Se  computing time  $\sim$
21 days  for both parity.  The calculations were performed with shell-model code ANTOINE \cite{Antoine}.



\section{Spectra analysis}
The results for the three interactions for  different model spaces used in the calculations
are presented with respect to the experiment in Figs.2-5.  
Earlier shell model result in
28-50 model space for pairing plus quadrupole-quadrupole interaction
has been reported in literature by Yoshinaga {\it et al.} \cite{yoshi}.  Further this work 
will add more information to the earlier work \cite{yoshi}, by
including $f_{7/2}$ orbital in the model space to study the proton
excitation across $Z=28$ shell. The shell model structure of Se isotopes was  discussed in~\cite{Honma09} using JUN45 interaction.
The results for the three interactions used in the calculations
are presented with respect to the experiment.  \label{E_BE}

\subsection{$^{78}{\rm Se}$}

\begin{figure*}
\begin{center}
\resizebox{1.03\textwidth}{!}{
\includegraphics{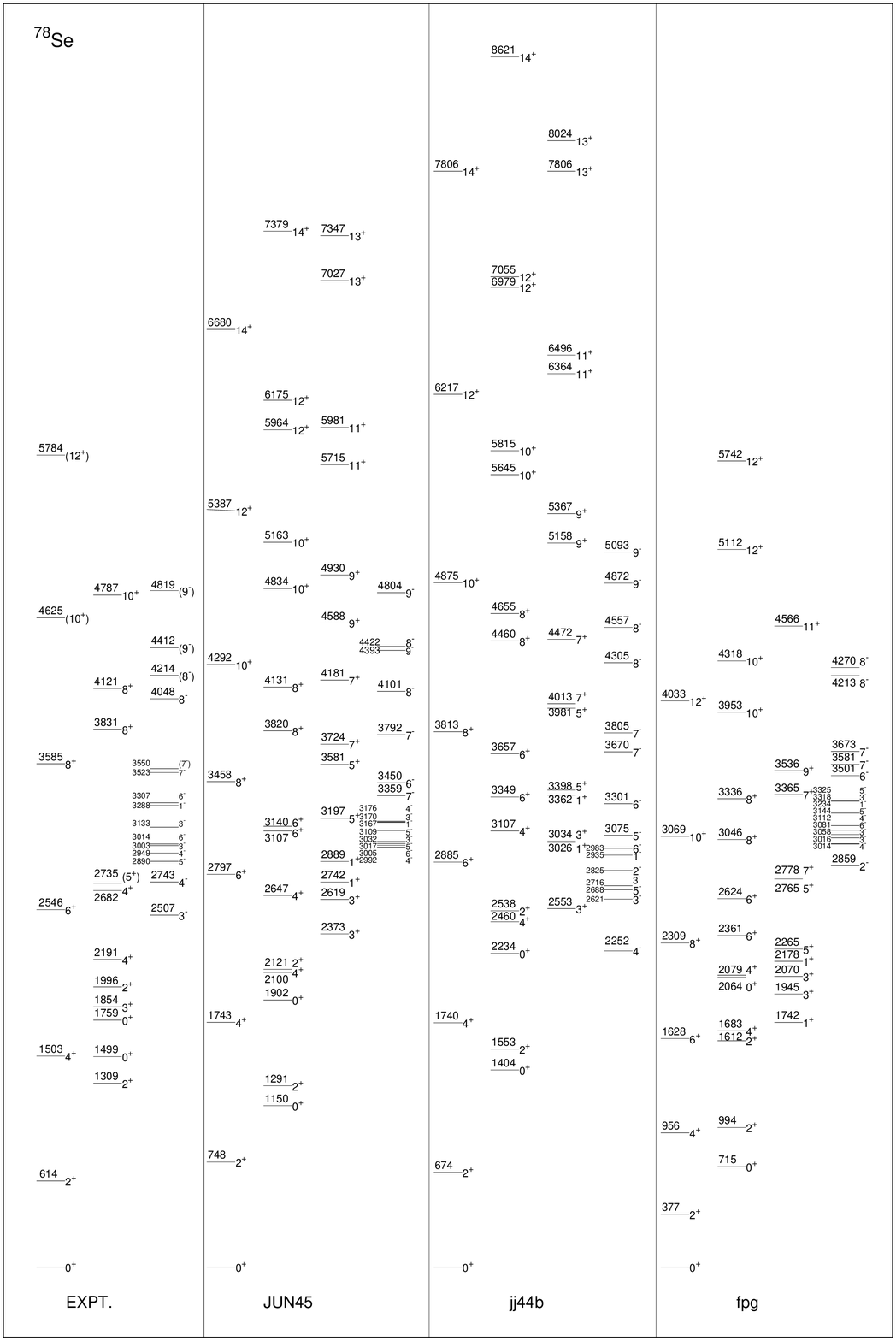} 
}
\caption{\label{f_se78}Comparison of experimental and calculated excitation spectra of $^{78}$Se with three different interactions.}

\end{center}
\end{figure*}

Comparison of the calculated values of the energy levels of $^{78}$Se with the experimental data is shown in  Fig.~\ref{f_se78}.
All the  three interactions correctly reproduces the experimental sequence of $0_1^+$, $2_1^+$, $4_1^+$, $6_1^+$, $8_1^+$, $10_1^+$ and $12_1^+$
levels at 614, 1503, 2546, 3585, 4625 and 5784 keV, respectively. In case of JUN45, the
calculated  $2_1^+$, $4_1^+$ and $6_1^+$ levels are 134, 240 and 251 keV higher than the experimental ones, while remained $8_1^+$, $10_1^+$ and $12_1^+$ are predicted 117, 323 and 397 keV lower than experimental levels. The $2_1^+$ level calculated by jj44b is only 60 keV higher as compared to its experimental counterpart. Difference between experimental and calculated energies is increased by increasing the excited energies reaching 433 keV at $12_1^+$. All the levels calculated by $fpg$ interaction is much lower than in the experiment. By the increasing spin the difference between the calculated and experimental levels is increased and it become rather 1751 keV when reaching $12_1^+$ level. 

In all the calculations first two $0^+$ and $2^+$ levels of the second positive-parity band are interchanged with respect to  experimental ones. The next $0^+$ and $3^+$ levels are located higher than in the experiment in all calculations. The experimental sequence levels $2^+$, $4^+$, $4^+$ and $5^+$ at 1996,  2191, 2682 and 2735 keV are better reproduced by the JUN45 calculation. The two $6^+$ levels appear in all calculations after the $5^+$ level which have not been observed in the experiment. The last three levels of this band calculated by the JUN45 interaction excellently agree with the experimental data. Experimental levels of this band measured up to $10^+$ and many odd spin levels $1^+$, $7^+$, $9^+$, which appear in the calculations, have not been observed in the experiment. 
  With the both the interactions the most dominant configuration for g.s. is $\pi (p_{3/2}^2f_{5/2}^4)$$\otimes$$\nu (p_{3/2}^4f_{5/2}^4p_{1/2}^2
g_{9/2}^6)$. The probability is less than 10\% in all wave function.
   
The experimental negative-parity levels start with the $3^-$ level. The first negative-parity level is $4^-$ in both JUN45 and jj44b calculations, while it is $2^-$ in the $fpg$ calculation. In the JUN45 calculation the first few levels are much compressed than in the experiment. The  spins and parities of $(7^-)$, $(8^-)$, $(9^-)$, and $(9^-)$ levels which is not confirmed in the experiment, are predicted by all calculations as negative parity levels.
The sequence of experimentally observed levels starting from $1^-$ is exactly the same as in the jj44b calculation.
 The structure of $3_1^-$ is $\nu (p_{1/2}
g_{9/2}^{-1})$ with 13\% for jj44b.

\subsection{$^{80}{\rm Se}$}

\begin{figure*}
\begin{center}
\resizebox{1.03\textwidth}{!}{
\includegraphics{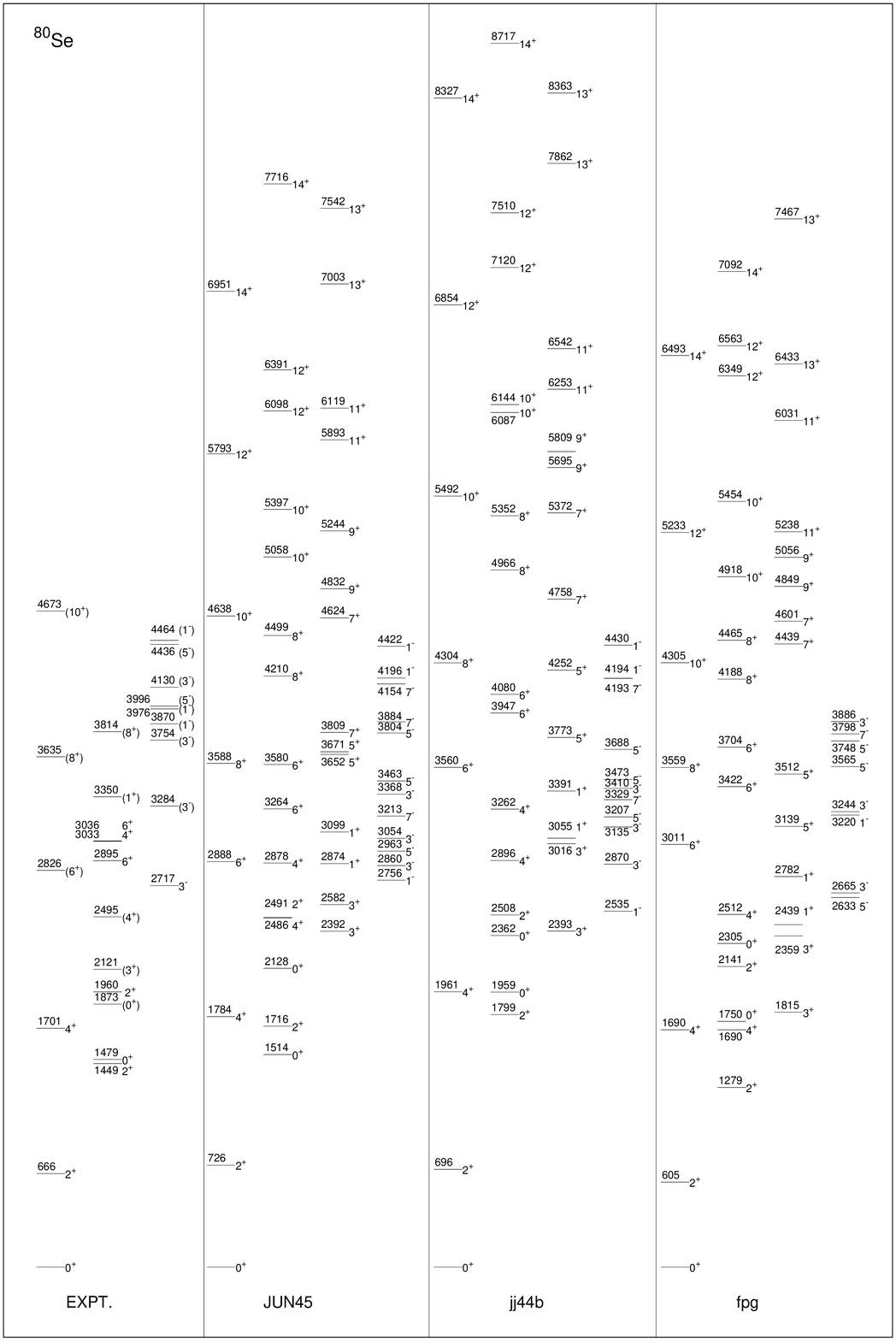} 
}
\caption{\label{f_se80}Comparison of experimental and calculated excitation spectra of $^{80}$Se with three different interactions.}
\end{center}
\end{figure*}

In  Fig.~\ref{f_se80} we have shown the comparison of the values of the energy levels calculated by
JUN45, jj44b and $fpg$ interactions with experimental data. The values calculated  by JUN45 are in
very good agreement with the all levels of the first experimental positive-parity band which are
available up to $10^+$. The jj44b calculation predicts higher values after $4^+$, while the values
predicted by the $fpg$ calculation are lower up to $4^+$ and then the $6^+$ level is higher.
The $8^+$ and $10^+$ levels are again lower than in the experiment.

The fist two levels of the second positive-parity band are interchanged with respect to those
of $^{78}$Se in jj44b calculation. The sequence of these levels are not changed with respect
to those of $^{78}$Se in JUN45 calculation.   In the $fpg$ calculation $2^+$ level is lowest
in the second positive-parity band. The $4^+$ levels is predicted much lower than in the
experiment, while the second $0^+$ and $2^+$ levels  are higher in this band than in the
experiment. The second $4^+$ level is only 17 keV higher than in the experiment. Then three
$6^+$ levels in succession come in the calculation.  The $8^+$ in the calculation is much
higher than in the experiment. For odd spin-positive-parities, which we have shown in the
third column, all the calculation predict two levels for each spin ($1^+$ and $1^+$, $3^+$
and $3^+$,..). In the experiment only one $1^+$ and $3^+$ are measured, first of which is
higher than all $1^+$ levels in all calculation, while second of them lower than $3^+_1$ level
in all calculations. For $^{80}$Se with both interaction probability $\sim$ 15\% but JUN45 have structure 
$\pi (p_{3/2}^2f_{5/2}^4)$$\otimes$$\nu (p_{3/2}^4f_{5/2}^6p_{1/2}^2
g_{9/2}^6)$ and jj44b have $\pi (p_{3/2}^2f_{5/2}^4)$$\otimes$$\nu (p_{3/2}^4f_{5/2}^4p_{1/2}^2
g_{9/2}^8)$.

The lowest negative-parity level according to JUN45 and jj44b  calculations are $1^-$, which
is much higher in the experiment, however the second and 
the third $1^-$ levels are closer to the experimental ones in both calculations. Four $3^-$ levels
are measured in the experiment. For both JUN45 and jj44b calculations the reported three $3^-$
levels are located approximately the same in both calculations. We have also reported three $5^-$
levels in all calculations. All of them are lower than experimentally measured two $5^-$ levels.

\subsection{$^{82}{\rm Se}$}

\begin{figure*}
\begin{center}
\resizebox{1.03\textwidth}{!}{
\includegraphics{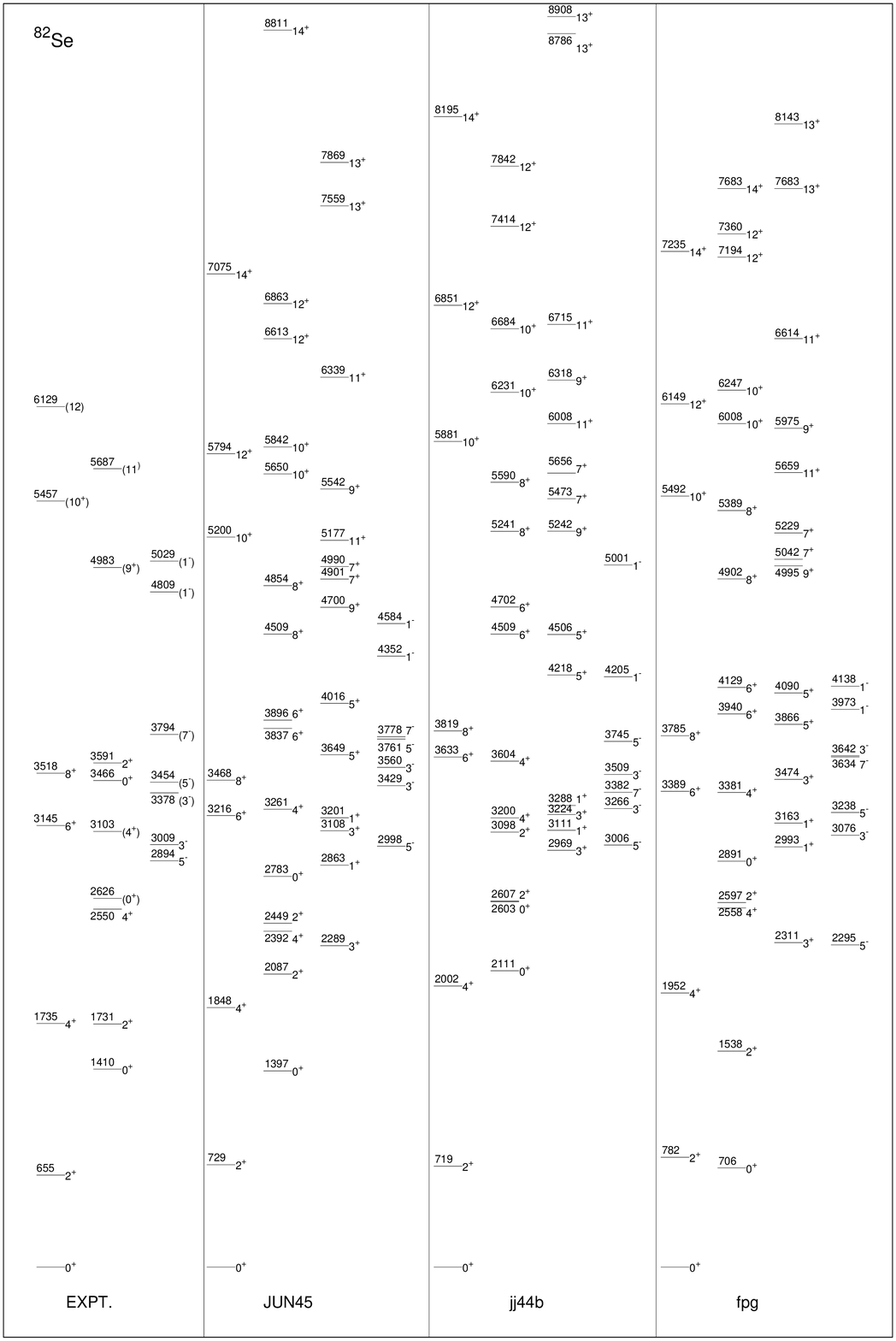} 
}
\caption{\label{f_se82}Comparison of experimental \cite{Porquet} and calculated excitation spectra of $^{82}$Se with three different interactions.}
\end{center}
\end{figure*}

As is seen from Figure~\ref{f_se82} in all calculated levels in the first column are in similar 
pattern to experimental one. Better values are predicted by JUN45 calculation. 

In all calculations the first level in the second column is $0^+$ level like in the experiment.
Closest value of this level is predicted by JUN45. The spacing between the first and the second
$2^+$ levels in the second column are less than that of in the experiment in all calculations.
In JUN45 calculation first of two $4^+$ levels is located lower while the second one is higher
than in the experiment. In jj44b calculation both of them are much higher than in the experiment.
In $fpg$ calculation the first $4^+$ is only 8 keV higher than in the experiment. The second $0^+$
is predicted well by jj44b calculation.

Among the positive-parity odd spin levels only $9^+$ and  $(11)$ level energies measured in
the experiment. The $(11)$ level's parity is predicted to be positive by the calculations.
As in the case of previous isotopes there are many other $1^+$, $3^+$, $5^+$ levels in the
calculation which are not measured in the experiment.

As is seen from the Figure~\ref{f_se82} agreement of the all calculated values of negative-parity
levels with the experimental data is improved very much as compared to the $^{78,80}$Se isotopes.

 For $^{82}$Se with JUN45 have structure 
$\pi (p_{3/2}^2f_{5/2}^4)$$\otimes$$\nu (p_{3/2}^4f_{5/2}^6p_{1/2}^2
g_{9/2}^8)$ and jj44b have $\pi (p_{3/2}^2f_{5/2}^4)$$\otimes$$\nu (p_{3/2}^4f_{5/2}^6p_{1/2}^2
g_{9/2}^8)$ with probability 39\% and 35\%, respectively.

\subsection{$^{84}${\rm Se}}
\begin{figure*}
\begin{center}
\resizebox{1.03\textwidth}{!}{
\includegraphics{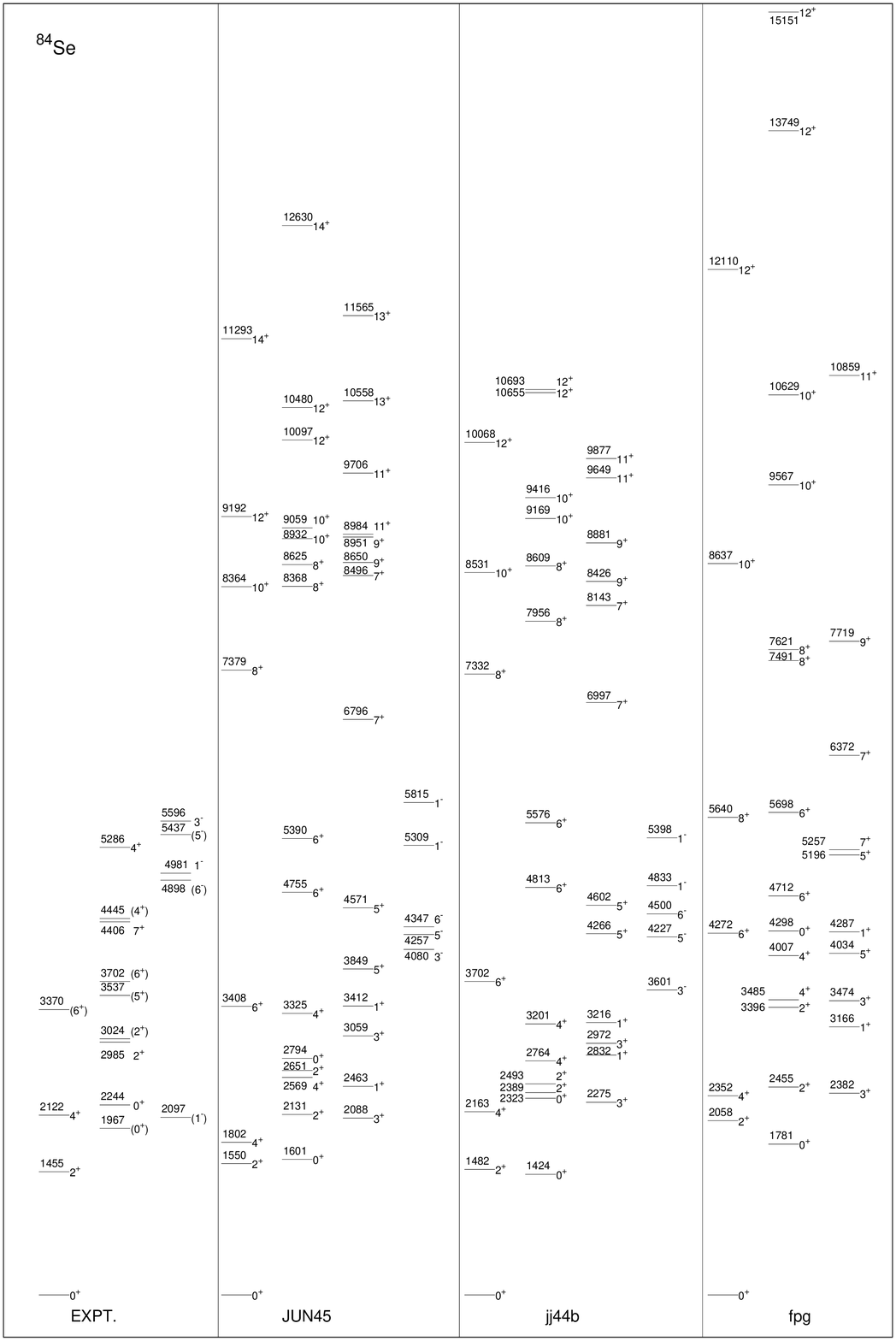} 
}
\caption{\label{f_se84}Comparison of experimental \cite{Gadea,Prevost,Jones} and calculated excitation spectra of $^{84}$Se with three different interactions.}
\end{center}
\end{figure*}

The levels of the first column of Figure~\ref{f_se84} are predicted better by jj44b calculation.
Also $6^+$ level is little bit higher than the in JUN45, spacings between the levels very much like
to the experimental ones. 

In all calculations the first positive-parity level in the second column is $0^+$ as in the experiment.
The sequences of $0^+_1$, $0_2^+$ and $2^+_1$, $2^+_2$ levels is the same with the experimental one in
jj44b calculation and spacings between them are larger than in the experiment. Sequence of the pair of
these levels are different from the experimental one in JUN45 and $fpg$ calculations and spacing between
them are much larger than in the experiment. The $4^+_1$, and $4^+_2$  levels are located much lower
than in the experiment in all calculations.  For ground state all the orbitals are completely filled with maximum probability 
of 40\%. The structure of $0_2^+$ is  $\pi (f_{5/2}^6)$$\otimes$$\nu (p_{3/2}^4f_{5/2}^6p_{1/2}^2
g_{9/2}^{10})$ with probability 33\% (JUN45) and 57\% (jj44b).

The $1^+$ and $3^+$ levels appearing in the calculations have not been measured in the experiment.
The measured $5^+$ level is predicted better by JUN45. The $7^+$ level is located very high in all
the calculations as compared to experimental one. 

The structure of the negative-parity levels for $^{84}$Se is changed so drastically as compared
to previous isotopes and both JUN45 and jj44b calculations fail to explain it.

For $^{78}$Se in Ref. \cite{yoshi}, the predicted $0^+_2$ state lies lower in energy compared to experimental data,
while in present work with jj44b interaction the result is close to experimental data with difference of only 95 keV.
In case of $^{82}$Se the predicted energy gap between the $6^+_1$ and $8^+_1$ states correctly reproduced while in 
Ref. \cite{yoshi} these levels are very close to each other.

\section{\label{ep} Transition probability, quadrupole moments and occupation numbers analysis}
\subsection{\label{ep} $E2$ transition probability and quadrupole moments}

The calculated $B(E2)$ transitions are shown in table 1.
In case of $^{78,80}$Se, the results predicted by $fpg$ interaction show better agreement
with experimental data. While for $^{82,84}$Se, the jj44b interaction results are more
reasonable. Thus we may conclude that proton excitation across $Z=28$  shell for
lighter Se isotopes are important.  In the present study our predicted result for $B(E2)$ values are close to
experimental data in comparison to Ref. \cite{yoshi}. The results of $B(E2)$ transitions for $fpg$ interaction
are probably related with the contribution of the protons in
the $f_{7/2}$ orbital, which could be compensated with smaller effective charges.  However with JUN45 and jj44b
due to missing proton $f_{7/2}$ orbital we need higher value of effective charges. To see the effect of inclusion
of proton $f_{7/2}$ orbital in the model space, we keep effective charges same. 

We have also calculated static quadrupole moments as shown in table 2. For the first 2$^+$ state,
the JUN45 and jj44b interaction correctly predict sign of quadrupole moments for $^{78,80,82}$Se.
The $fpg$ interaction predicts positive sign for $^{78}$Se. The overall
agreement for quadrupole moments predicted by jj44b is better.

\begin{table}[htbp]
\begin{center}
\caption{$B(E2)$ reduced transition strength in W.u. Effective charges
  $e_p=1.5$ $e_n=0.5$ were used. Experimental values were taken from
  the NNDC database.}
\vspace{0.2cm}
\label{tab:table2}
\begin{tabular}{ c | c | c | c | c } \hline 
 & $^{78}$Se&$^{80}$Se&$^{82}$Se& $^{84}$Se \\ \hline
BE($2_1^+ \rightarrow 0_1^+$)  && & &   \\ \hline
Experiment & 32.8(4.5) & 24.2 (0.4) & 16.7  (0.3) & N/A \\ 
JUN45  & 18.83 & 16.53 & 13.88  & 6.64 \\ 
jj44b  & 20.74 & 18.88 & 15.33 & 8.37 \\ 
$fpg$  & 30.87 & 24.08  &16.33 & 8.22   \\ 
 && & &  \\ \hline
BE($4_1^+ \rightarrow 2_1^+$)  && & &\\ \hline
Experiment & 38.2$^{+5.6}_{-5.1}$ &34.7 (1.1) & 18.7 (3.0) & N/A  \\ 
JUN45 & 25.47 & 21.40 & 19.60  & 1.96\\ 
jj44b & 27.23& 25.65 & 20.77  & 0.005 \\ 
$fpg$ & 46.27  & 32.85  &  25.86  & 1.42  \\ 
 && & &  \\ \hline
BE($6_1^+ \rightarrow 4_1^+$) & & & &\\ \hline
Experiment & 48(14) & N/A  & N/A  & N/A  \\ 
JUN45  & 23.69 & 16.81 & 15.42  & 1.02 \\ 
jj44b  & 24.22 & 21.92 & 18.01  & 3.25 \\ 
$fpg$  & 48.92  &  32.01 & 24.42  &  0.12 \\ 
 & & & &  \\ \hline
BE($8_1^+ \rightarrow 6_1^+$)  && & & \\ \hline
Experiment  & 57(19) & N/A & 0.56 (0.03)   & N/A \\ 
JUN45   & 13.01& 7.45 & 0.38  & 0.0008 \\ 
jj44b   & 21.86& 1.27 & 0.31  & 0.002 \\ 
$fpg$    & 49.01 & 7.13  &  0.57  & 1.13  \\ 
& & & &  \\ \hline 
\end{tabular}
\end{center}
\end{table}

\begin{table}[htbp]
\begin{center}
\caption{ Electric quadrupole moments, $Q_s$ (in eb), with the three different interactions (the effective
charges $e_p$=1.5, $e_n$=0.5 are used in the calculation).}
\vspace{2mm}
\label{tab:table3}
\begin{tabular}{ c | c | c | c | c } \hline
&$^{78}$Se &$^{80}$Se&$^{82}$Se& $^{84}$Se \\ \hline
Q($2_1^+$)   && &  &  \\ \hline
Experiment & -0.20  (7) & -0.31  (7) & -0.22 (7)  & N/A  \\ 
JUN45  &-0.13 &-0.31 & -0.33  & +0.01 \\ 
jj44b &-0.32 & -0.36 & -0.37  & -0.27 \\ 
$fpg$  &+0.47 &-0.35 & -0.36    & +0.04  \\ 
       & & & &  \\ \hline
Q($2_2^+$)  && &  &  \\ \hline
Experiment & +0.17  (9) & N/A &N/A  &N/A  \\ 
JUN45  &+0.13& +0.28 & +0.24  & -0.09 \\ 
jj44b &+0.30&+0.35 & +0.27  & +0.13 \\ 
$fpg$  &-0.33 &+0.36 &+0.28    & +0.007  \\ 
        & & &  \\ \hline
Q($4_1^+$)  & & &  &  \\ \hline
Experiment  & -0.68 (15) &N/A &N/A  &N/A  \\ 
JUN45 &-0.09 &-0.35&-0.39  & +0.10 \\ 
jj44b &-0.36 &-0.40&-0.42  & +0.18 \\ 
$fpg$ &+0.63 &-0.29 & -0.43    & +0.15  \\ 
        & & &  \\ \hline
\end{tabular}
\end{center}
\end{table}

\subsection{\label{on} Occupation numbers}

In Fig. 6, we show the proton/neutron occupation numbers of
$fpg$-shell orbits for $0_1^+$ and $2_1^+$ levels. The proton
occupancies increasing smoothly in the $f_{5/2}$ orbital, while
$p_{3/2}$ decreasing. But beyond $N=46$, in case of JUN45
 interaction the change in occupancy of these two orbitals are
very significant.  With both interaction the occupancy of proton  $p_{1/2}$ and $g_{9/2}$
orbitals are similar. As the neutron number increases the occupation
number of $\nu g_{9/2}$ orbital increases drastically. The neutron
number occupation show a similar distribution for $2_1^+$ as in the
ground state.

\newpage
\begin{figure*}
\begin{center}
\vskip -2.cm
\resizebox{1.1\textwidth}{!}{
\includegraphics{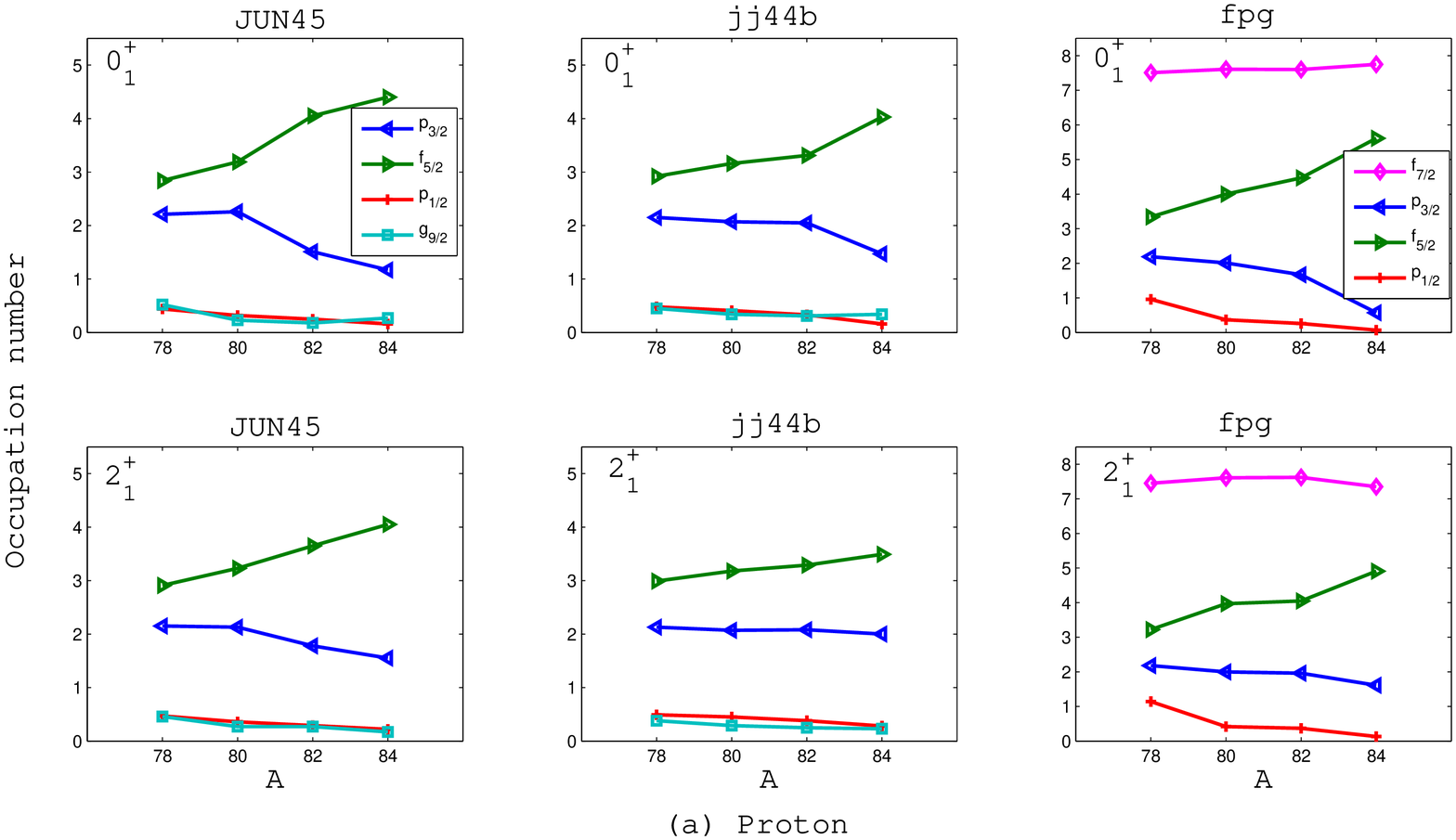} 
}
\resizebox{1.1\textwidth}{!}{
\includegraphics{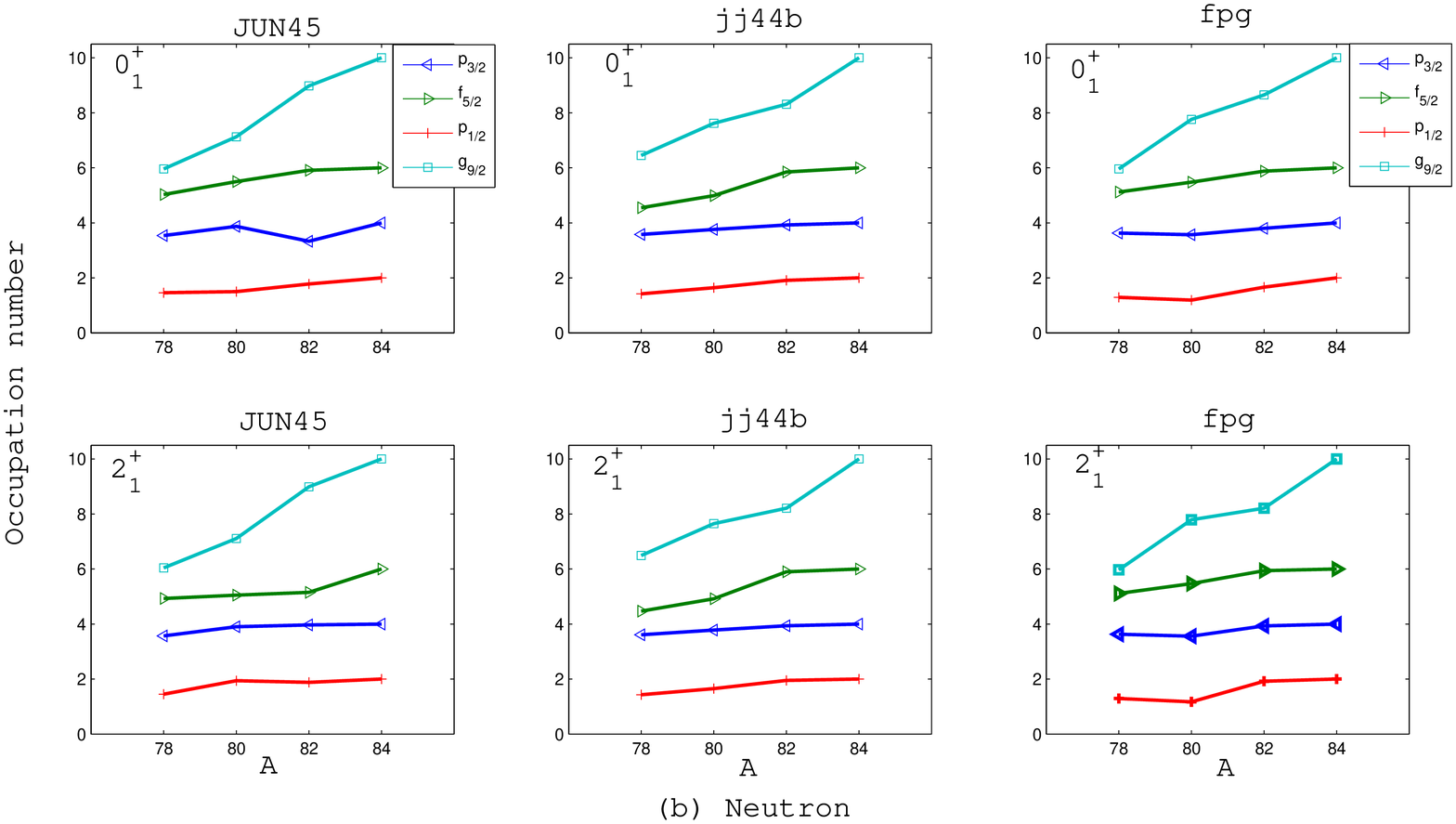} 
}
\caption{\label{Fig10}(Color online) Proton/Neutron occupation numbers of the JUN45 and jj44b  ($p_{3/2}$, $f_{5/2}$, $p_{1/2}$ and $g_{9/2}$ -shell orbits) and $fpg$ ($f_{7/2}$, $p_{3/2}$, $f_{5/2}$, $p_{1/2}$ -shell orbits) interactions-
 for two low-lying states in even-even Se isotopes. (Upper
panel) $0_1^+$ states;(lower panel) $2_1^+$ states.
}
\end{center}
\label{f_82ge}
\end{figure*}


\section{\label{sec5}Summary}
\label{conc}

In summary, comprehensive study for the structure of
neutron-rich even-even Se isotopes have been carried out using large-scale shell-model calculations for
two spaces: full $f_{5/2}pg_{9/2}$ space and $fpg_{9/2}$ space with
$^{48}$Ca core. \\

The following broad conclusions can be drawn: 

\begin{itemize}  

\item The overall calculated results for the energy levels,
$B(E2)$s and quadrupole moments are in good agreement with the
experimental data.

\item  The $E2$ transitions, quadrupole moments analysis
show the importance of proton excitations across $Z=28$ shell for
 $fpg_{9/2}$ space. 

\item  For $^{78-80}$Se the $B(E2)$ values predicted by $fpg$
  transitions are in better agreement with experimental data.

\item  The result (wave functions) of $^{82}$Se, may be used 
for calculating nuclear transition matrix elements and finally
half-live for this nucleus which is a good candidate for neutrinoless
double beta decay.

\item  It is also important while tuning the effective interaction 
we can also take experimentally known quadrupole moment as
a parameter to increase predictive power of effective interaction.

\item  Further theoretical development is needed by enlarging model
  space by including $\nu d_{5/2}$ orbital to study simultaneously
  proton and neutron excitations across $Z=28$  and $Z=50$ shell,
respectively.

\end{itemize} 

\section*{Acknowledgments} 
Thanks are due to
E. Padilla - Rodal for useful discussions during this work.
All the shell-model calculations have been performed at KanBalam
computational facility of DGCTIC-UNAM, Mexico. 
MJE acknowledges support from grant No. 17901 of
CONACyT projects CB2010/155633 and F2-FA-F177 of Uzbekistan Academy of
Sciences.

\section*{References}
\bibliographystyle{jphysg}

\begin{thebibliography}{10}
\providecommand{\url}[1]{\texttt{#1}}
\providecommand{\urlprefix}{URL }
\providecommand{\eprint}[2][]{\url{#2}}

\bibitem{Padalia}
Padilla-Rodal E  {\it et al} 2005
Phys.\ Rev.\ Lett. {\bf94} 122501 

\bibitem{Heyde13}
Heyde K 2013
Phys. Scr. {\bf T152} 014006 

\bibitem{Otsuka13}
Otsuka T 2013
Phys. Scr. {\bf T152} 014007

\bibitem{flangan09}
Flanagan K T  {\it et al} 2009
Phys.\ Rev.\ Lett. {\bf103} 142501 

\bibitem{cheal10}
Cheal B {\it et al} 2010
Phys.\ Rev.\ Lett. {\bf104} 252502

\bibitem{astier11}
Porquet  M -G {\it et al} 2011
Phys.\ Rev.\ C {\bf84} 054305

\bibitem{pcs_ga} 
Srivastava P C 2012
J.\ Phys.\ G\ {\bf39}  015102 

\bibitem{Honma09}
Honma M, Otsuka T, Mizusaki T and Hjorth-Jensen M 2009
Phys.\ Rev.\ C {\bf80}  064323

\bibitem{brown}
B.A. Brown and A.F. Lisetskiy (unpublished).

\bibitem{fpg}
Sorlin O {\it et al}  2002
Phys.\ Rev.\ Lett. {\bf88} 092501

\bibitem{32} A. Poves, J. Sanchez-Solano, E. Caurier, and
F. Nowacki, Nucl. Phys. A {\bf88}, 157 (2001).

\bibitem{33}  F. Nowacki, PhD Thesis (IReS, Strasbourg, 1996).

\bibitem{34} S. Kahana, H. C. Lee, and C. K. Scott, Phys. Rev.
{\bf180}, 956 (1969).

\bibitem{Antoine}
Caurier E,  Mart\'inez-Pinedo G, Nowacki F, Poves A, and Zuker A P 2005
Rev.\ Mod.\ Phys. {\bf77} 427 

\bibitem{yoshi}
Yoshinaga N, Higashiyama K, and Regan P H, 
Phys.\ Rev.\ C {\bf78}  044320 

\bibitem{Porquet} 
Porquet M -G {\it et al} 2009
Eur.\ Phys.\ J. \ A. {\bf39}  295 

\bibitem{Gadea} 
Gade A {\it et al} 2010
Phys.\ Rev.\ C {\bf81} 064326 

\bibitem{Prevost} 
Pr\'evost  A {\it et al } 2004
Eur.\ Phys.\ J. \ A. {\bf22}  391 

\bibitem{Jones} 
Jones E F  {\it et al} 2006
Phys.\ Rev.\ C {\bf73} 017301 





\end{thebibliography}

\end{document}